\documentclass{mem}
\usepackage{natbib}\usepackage{txfonts}\usepackage{balance}
\usepackage{graphicx}
\usepackage[a4paper]{hyperref}
\idline{0}{1}
\begin{document}
\def\teff{$T\rm_{eff }$}
\def\kms{$\mathrm {km s}^{-1}$}
\def\mdot{$\mathrm {M_{\odot} yr}^{-1}$}

\title{
Sh2-188: a model for a speedy PN
}

   \subtitle{}

\author{
C.J. \,Wareing\inst{1}, 
T.J. \,O'Brien\inst{1},
A.A. \,Zijlstra\inst{2}
\and J.E. \,Drew\inst{3}
          }

  \offprints{C.J. Wareing}

\institute{
Jodrell Bank Observatory,
School of Physics and Astronomy,
The University of Manchester,
Macclesfield, Cheshire,
SK11 9DL, United Kingdom,
\email{cwareing@jb.man.ac.uk}
\and
School of Physics and Astronomy,
The University of Manchester,
Oxford Road, Manchester,
M13 9PL, United Kingdom
\and
Imperial College of Science, Technology and Medicine,
Blackett Laboratory,
Prince Consort Road,
London,
SW7 2BW, United Kingdom
}

\authorrunning{Wareing et al.}

\titlerunning{Sh2-188}

\abstract{
Sh2-188 is thought to be an ancient planetary nebula in the galactic disk. It appears to be one-sided with recent observations revealing structure behind the filamentary limb. We postulate that Sh2-188 is interacting with the ISM and simulate it in terms of a ``triple-wind'' model comprising of the usual ``fast'' and ``slow'' interacting stellar winds plus the wind due to motion through the ISM. We have run simulations at various velocities of the central star relative to the ISM and find that a high velocity of 125 \kms \  best approximates the observed structure. We also suggest that Sh2-188 is younger than previously thought and that much of the mass lost on the AGB has been swept downstream.
}
\maketitle{}

\section{Introduction}

The planetary nebula Sh2-188 \citep{sharpless59} has a one-sided, semi-circular and filamentary appearance. It is located in the Galactic plane at $l = 128^\circ$, $b = -4^\circ$. Spectroscopic work \citep{parker64,lozinskaya71,johnson75} revealed an extremely high NII/H$\alpha$ line-ratio, thought to be due to an overabundance of nitrogen. Further spectroscopic work \citep{rosado82} found abundances similar to Peimbert Type I planetary nebulae (PN) \citep{peimbert81}. \cite{kwitter88} presented an identification of a candidate central star of the nebula using geometric, colour and apparent magnitude methods. The candidate star is not at the geometric centre of the optical filaments.

The distance to the candidate star has been estimated at $d = 965^{+1000}_{-600}$ pc by non-LTE modelling \cite{napiwotzki99,napiwotzki01} which also estimated its temperature to be $102000 \pm 32000$ K. Consistently, this nebula is found to be an evolved object. \cite{tweedy96} published an atlas of ancient PNe including Sh2-188. 

\begin{figure}[t!]
\begin{center}
{\includegraphics[clip=true,width=6.5cm]{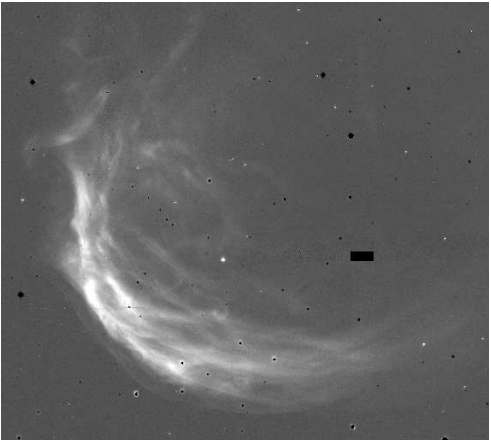}}
\caption{\footnotesize An image of Sh2-188 created by a mosaic of images taken as part of the IPHAS survey in 2003 (see text). North up and East to the left. The image is $9.4 \times 8.5$ arcminutes.}
\label{iphas-bow}
\end{center}
\end{figure}

\begin{figure}[t!]
\begin{center}
{\includegraphics[clip=true,width=6.5cm]{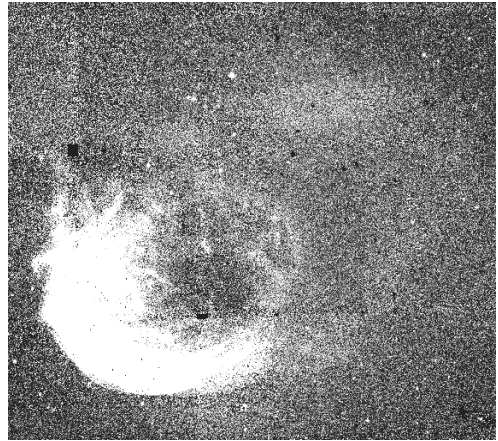}}
\caption{\footnotesize Same as Figure \ref{iphas-bow} but with scaling chosen to reveal further structure. North up and East to the left. The image shows a faint ring completing the bright semi-circular arc and tails stretching away and connecting in the north-east. The image is $18.8 \times 17.0$ arcminutes.}
\label{iphas-deep}
\end{center}
\end{figure}

Figure \ref{iphas-bow} shows an image of Sh2-188 created by a mosaic of H$\alpha$ observations taken as part of the Isaac Newton Telescope Photometric H$\alpha$ Survey of the Northern Galactic Plane (IPHAS) \citep{drew05}. The bright portion of the nebula in the southeast describes a semi-circular arc with a diameter of 10 arcminutes. Structure first noted by \cite{tweedy96} to the northwest of the nebula has now been revealed by the observations of \cite{drew05} to be a faint circular closure of the bright nebula with connecting tails stretching behind. Figure \ref{iphas-deep} shows this structure. From the front of the filamentary structure to the connecting tails the nebula is 15 arcminutes long. The tail seems to be stretching away from the nebula in opposition to an inferred direction of motion from the bright arc which could explain the one-sided brightening as an interaction with the ISM.

The accepted theory of PN formation is the interacting stellar winds model (ISW) \citep{kwok82} where a fast wind ($\sim10^3$ \kms) from the hot central star of a PN blows into the slow wind ($\sim10$ \kms) produced during the preceding AGB phase resulting in the inner regions of the slow wind being compressed into a dense shell and ionised by the energetic UV radiation of the central star. The familiar ring-like appearance of PNe is then observed. Interaction with the ISM was thought to be something that was only observed at late times \citep{tweedy96}. \cite{villaver03} (hereafter referred to as VGM) reopened the PN-ISM interaction debate by performing simulations which follow the full evolution of the AGB phase followed by the PN phase with a conservative proper motion of the central star. Their results showed that the AGB wind structure is strongly influenced by the ISM interaction and thus changes the appearance of the PN at early as well as late times.

Sh2-188 appears to be one of the most extreme cases of PN-ISM interaction, made particularly rare by its filamentary nature. We have developed a ``triple-wind'' model adding a third wind due to movement through the ISM into the ISW model. We are investigating whether this triple-wind model can reproduce the nebular shape of Sh2-188 using a new code developed by Wareing and O'Brien \citep{wareing05}.

\section{The numerical method}

The numerical scheme used to solve the hydrodynamic equations uses the second-order Godunov scheme due to \cite{falle91}. The scheme is in 3D cartesian coordinates, fully parallel and includes the effect of radiative cooling. A small amount of viscosity is required in order to control the effects of the Quirk instability \citep{quirk92}. The correction takes the form of an addition to the solution of the Riemann problem. This additional flux is small in regions of smooth flow and affects only regions containing discontinuities.

The numerical domain consists of a grid with 200 cells in each direction and uniform grid spacing. The central mass-losing star is placed at particular coordinates ($x_0,y_0,z_0$) and simulations are performed in the frame of reference of the star. Mass-loss is effected by means of artificially setting the values of density, momentum and energy density in the cells in a volume-weighted spherical region of radius 5$\frac{3}{4}$ cells centred on the star. The conditions within this region are reset at the beginning of every computational timestep so as to keep driving the wind. Movement through the ISM is set to be parallel to the $x$-axis and material flows in at the ($x=1,y,z$) boundary face with a positive $v_{\rm x}$. All other boundaries have free-flow conditions allowing material to flow out of the domain freely.

\section{The input parameters}

The ISW model of the PN-ISM interaction follows the evolution of the AGB and post-AGB phases from the beginning of the AGB phase through the post-AGB or PN phase. In the triple-wind model, this evolution is combined with a constant movement through the ISM. The evolution is followed in the frame of the central star emitting the stellar winds as the ISM flows through the computational grid with particular velocity $v_{\rm ISM}$.

Mass-loss via a stellar wind has been modelled as a spherically symmetric constant mass-loss rate $\dot{M}$ with constant velocity $v$. The density in the stellar winds is set according to $\rho \propto r^\alpha$. In the AGB and post-AGB winds $\alpha = 2$ and the constant of proportionality is $\dot{M} / 4\ \pi\ v$. In the ISM wind $\alpha = 0$ and the density and velocity are constant.

Typical estimates from the literature are used for the wind inputs. The fast stellar wind parameters are denoted by a subscript fw: $\dot{M}_{\rm fw} = 5 \times 10^{-8}$ \mdot, $v_{\rm fw} = 1000$ \kms \& $T_{\rm fw} = 5 \times 10^4$\ K. The slow stellar wind parameters are denoted by a subscript sw: $\dot{M}_{\rm sw} = 10^{-6}$ \mdot, $v_{\rm sw} = 10$ \kms \& $T_{\rm sw} = 10^4$\ K. The switch between the AGB wind and the post-AGB wind is instantaneous and occurs after $10^5$ years of AGB evolution, which is short but allows a stable structure to develop. The detailed properties of these winds, whilst becoming more certain, are still too uncertain to justify more detailed temporal modelling. We are aiming at developing a simple model to investigate the ISM interaction. The ISM itself is assumed to be homogeneous with characteristics of warm neutral medium, the main constituent of the observed ISM: $T_{\rm ISM} = 2500$\ K \& $\rho_{\rm ISM} = 1$\ cm$^{-3}$ \citep{burton88}. The gas pressure in all three winds is calculated ignoring magnetic pressure via $P = \rho\ k\ T\ / m_{\rm a}$ where $k$ is Boltzmann's constant and $m_{\rm a}$ is the average mass of a particle.

\section{Results}

Various values of $v_{\rm ISM}$ have been considered from 0 \kms\ to 175 \kms\ in steps of 25 \kms. The evolution of these models can be found in \cite{wareing05}. We present here only the case of 125 \kms\ which we find best reproduces the general morphology of Sh2-188.

\begin{figure}[t!]
\begin{center}
{\includegraphics[clip=true,width=6.5cm]{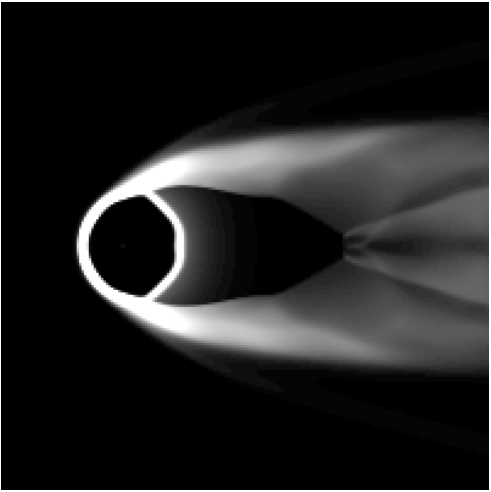}}
\caption{\footnotesize A slice through the computational grid at ($x,y = 100,z$) showing the density of the gas 2000 years after the onset of the fast post-AGB wind. The scaling is logarithmic.}
\label{simulation}
\end{center}
\end{figure}

Figure \ref{simulation} shows a density slice through the computational grid ($x,y = 100,z$) 2000 years after the onset of the fast wind. This snapshot has been selected for its resemblance to Sh2-188. The gas structure shows a strong bow-shock in the direction of motion, a faint shell of compressed AGB material completing the circle and connecting tails downstream of the main nebula made up purely of AGB material. Considering the evolution of the gas up to this point, the AGB wind has formed a bow-shock interaction upstream of the nebula. The compression and deformation has formed a high density bow-shock which reached a stable state between the material supplied by the AGB wind and ram-pressure stripping of AGB material downstream after $60,000$ of the $100,000$ years of AGB evolution. The material we see in the tail is purely AGB material and thus far older than the nebula. During the PN phase the wind decreases in density and increases in velocity resulting in an adiabatic shock that forms the bright shell of the nebula. Initially this shock is within the region of undisturbed AGB wind inside the inner shock and the nebula appears circular. After 2000 years, the fast adiabatic shock has expanded enough to interact with the bow-shock and strong asymmetries appear, similar to the appearance of Sh2-188. After this time, the nebula departs from a circular appearance as the adiabatic shock moves downstream through the undisturbed AGB material.

\section{Discussion}

We have found that the triple-wind model can reproduce the appearance of the nebula Sh2-188. If we accept the model, the nebula would appear to be moving at a high speed of $125$ \kms \  through the ISM. The typical velocity of a central star which results in a PN has been shown to be in the area of $35 - 50$ \kms \ \citep{binney98}. Thick disc stars resulting in a PN have a higher average velocity of $50 - 75$ \kms \ \citep{binney98}. The central star of Sh2-188 is therefore a rare object for it to be moving this quickly. Having investigated other proper motions and the resulting nebula shapes, a velocity of less than $75$ \kms\ does not result in connecting tails behind the nebula. At $100$ \kms\ fluid instabilities seem to destroy the smooth tail structure we observe and thus $125$ \kms\ is in fact the lowest speed at which we see a nebula resembling Sh2-188.

The general idea of a PN moving through the ISM was discussed in depth by \cite{dgani94,dgani98}. Rayleigh-Taylor instabilities are thought to be able to fragment a bow-shock in the direction of motion. Magnetic fields are shown to be able to alter the fragmentation. In the case of Sh2-188, it is thought to be in the galactic plane and thus the observed fragmentation can be interpreted \citep{dgani98} as being an effect of fluid and magnetic field instabilities. Importantly, \cite{dgani98} conclude that fragmentation of a substantial part of the halo requires a minimum proper motion of $100$ \kms. This limiting speed would predict that if Sh2-188 has a proper motion of $125$ \kms\ then the bow-shock would fragment. Whilst we do not see this in the simulations, we do observe this in the images e.g. Figure \ref{iphas-bow}. The lack of fragmentation in our simulations may be the result of a lack of resolution in the computational grid.

VGM pointed out that the problem in available models of PN-ISM interaction is that the interaction had only been studied by considering the relative movement when the nebular shell was already formed. The 2D simulations they performed for typical PN conditions (ISM density: $0.1$ cm$^{-3}$; proper motion: $20$ \kms) followed the full AGB and post-AGB evolution of the stellar wind. They found that high velocities, high ISM densities and magnetic fields, some or all of which had previously been required to produce observed ISM interactions, were no longer necessary to explain the observed asymmetries in the external shells of PN. Further, the stripping of mass downstream during the AGB phase provides a possible solution to the problem of missing mass in PN whereby only a small fraction of the mass emitted during the AGB phase is inferred to be present during the post-AGB phase. The shape and appearance of our simulated PN is considerably affected by the evolution of the stellar wind during the AGB phase. We have shown that at high proper motions, the ISM interaction is apparent within a few thousand years of the onset of the fast wind and is strongly influenced by the preceding AGB evolution.

There is now little doubt that interaction with the ISM considerably alters the amount of mass within the circumstellar envelope during the AGB and post-AGB phases. Much of the mass emitted from the central star is downstream and in fact forms some of the structures we observe. In our simulation we find that having injected $0.102$ M$_{\odot}$ of material into the nebula region, only $0.005$ M$_{\odot}$ remains and $0.096$ M$_{\odot}$ has been swept downstream. It is important to note that these results are for only one set of stellar wind and ISM parameters. Tuning the parameters can produce larger nebula radii which take longer to reach a point of stability and thus more material is in the nebula region. This also makes it difficult to place a constraint on the distance of Sh2-188.

The faint emission completing the observed ring is particularly important. In the simulations this structure has a transitory nature and moves quickly away from the central star. This faint ring may be a very good tracer for the age of the nebula. The presence of the ring close to the star and still appearing to complete the circle of the nebula leads us to suggest that Sh2-188 is of an intermediate age, not an ancient object.

\section{Conclusions}

We have simulated the formation of a PN where the central mass-losing star moves through the ISM and have found that we can reproduce the general shape of the nebula Sh2-188. Following the AGB evolution has been the only way to recreate the whole structure, in particular the tail behind the nebula comprising of AGB material alone. ISM interaction in PN is observable from an early to intermediate age; Sh2-188 is most likely of an intermediate age, indicated by the position of the transitory closed ring which rapidly evolves downstream at the start of the PN phase. The ISM interaction can explain the problem of missing mass in PN and simulations can quantify this effect. The next generation of telescopes, particularly ALMA, will reveal cool dust structure in the universe and shed light on the AGB material swept downstream of moving PN.

\begin{acknowledgements}

C. Wareing has been supported by a PPARC grant.

\end{acknowledgements}

\bibliographystyle{aa}

\end{document}